\documentclass[11pt]{article}
\usepackage{amsmath,amssymb,amsthm,amsrefs}

\newtheorem{theorem}{Theorem}

\newtheorem{lemma}[theorem]{Lemma}

\newcommand{\N}{{\mathbb{N}}}

\newcommand{\Z}{{\mathbb{Z}}}

\newcommand{\cC}{{\mathcal C}}

\newcommand{\cE}{{\cal E}}

\newcommand{\onemat}{I}

\newcommand{\tr}{\mathop{\mathrm{tr}}}
\newcommand{\rank}{\mathop{\mathrm{rank}}}
\newcommand{\poly}{\mathop{\mathrm{poly}}}
\newcommand{\E}{\mathop{\mbox{$\mathbf{E}$}}}

\renewcommand{\>}{\rangle}
\newcommand{\<}{\langle}
\newcommand{\ket}[1]{|#1\rangle}
\newcommand{\bra}[1]{\langle #1|}

\newcommand{\norm}[1]{\|#1\|}

\newcommand{\triv}{{\hat 1}}

\newcommand{\semidirect}{\rtimes}
\renewcommand{\setminus}{-}

\newcommand{\tgamma}{\tilde{\gamma}}

\newcommand{\be}{\begin{equation}}
\newcommand{\ee}{\end{equation}}
\def\ba#1\ea{\begin{align}#1\end{align}}

\begin{document}


\title{\Large \textbf{On the quantum hardness of solving isomorphism \\
problems as nonabelian hidden shift problems}} 

\author{
Andrew M.\ Childs\footnote{amchilds@caltech.edu}{~}
and
Pawe{\l} Wocjan\footnote{wocjan@cs.caltech.edu} \\[1ex]
Institute for Quantum Information \\
California Institute of Technology \\
Pasadena, CA 91125, USA
}

\date{}

\maketitle

\abstract{We consider an approach to deciding isomorphism of rigid
$n$-vertex graphs (and related isomorphism problems) by solving a
nonabelian hidden shift problem on a quantum computer using the
standard method.  Such an approach is arguably more natural than
viewing the problem as a hidden subgroup problem.
We prove that the hidden shift approach to rigid graph isomorphism is
hard in two senses.  First, we prove that $\Omega(n)$ copies of the
hidden shift states are necessary to solve the problem (whereas $O(n
\log n)$ copies are sufficient).  Second, we prove that if one is
restricted to single-register measurements, an exponential number of
hidden shift states are required.}

\section{Introduction}

One of the major challenges of quantum computing is to determine
whether there exists an efficient quantum algorithm to decide if two
graphs are isomorphic.  It is well known that the graph isomorphism
problem can be reduced to a hidden subgroup problem over the symmetric
group \cites{BL95,Bea97,Hoy97,EH99}.  This approach seems to be
promising since hidden subgroup problems over many groups, including
arbitrary abelian groups \cites{Sim97,Sho97,BL95,Kit95,HH00} and some
nonabelian ones \cites{IMS03,FIMSS03,GSVV04,Gav04,MRRS04,BCD05b} can
be solved efficiently on a quantum computer.  An efficient quantum
algorithm for graph isomorphism would be interesting since no
efficient classical algorithm for the problem is known; the best known
classical algorithm for deciding isomorphism of $n$-vertex graphs runs
in time $O(n^{\sqrt{cn/\log n}})$ for some constant $c$ \cite{BKL83}.

Unfortunately, the only results so far on the quantum complexity of the
graph isomorphism problem consist of evidence that the problem might
be hard (with the notable exception of the result that the query
complexity of the associated hidden subgroup problem is polynomial
\cite{EH99}).  The graph isomorphism problem can be reduced to a
hidden subgroup problem in $S_{2n}$ where the hidden subgroups are
generated by full support involutions.  Hallgren, Russell, and Ta-Shma
showed that {\em weak Fourier sampling}, in which one performs a
nonabelian Fourier transform but then only measures the name of an
irreducible representation, is insufficient to solve the problem
\cite{HRT03}.  Kempe and Shalev generalized their result to show that
finding other subgroups of the symmetric group is also hard
\cite{KS04}.  Finally, Moore, Russell, and Schulman have obtained
results about the need to use multi-register measurements on the
hidden subgroup states obtained by Fourier sampling.  In particular,
if one is restricted to single-register measurements (in the standard
approach known as {\em strong Fourier sampling}), an exponential
number of hidden subgroup states is required \cite{MRS05}.  Similarly,
if one is restricted to two-register measurements, then a
superpolynomial (though possibly subexponential) number of hidden
subgroup states is required \cite{MR05}.  Strictly speaking, these
results do not show that the hidden subgroup problem directly relevant
to graph isomorphism is hard, since the possible subgroups resulting
from the graph isomorphism reduction are not generated by arbitrary
full support involutions, but by involutions having further properties
(as we will discuss further in Section~\ref{sec:iso}, in connection
with the hidden subgroup problem over $S_n \wr \Z_2$).  However,
concurrently with the present work, Moore, Russell, and Schulman have
improved their result for single-register measurements to cover the
special case directly relevant to graph isomorphism \cite{MRS05comm}.

In this paper, we study an alternative approach to solving graph
isomorphism on a quantum computer, by viewing it as an instance of a
nonabelian hidden {\em shift} problem.  This approach is arguably more
natural than viewing the problem as a hidden subgroup problem: every
possible hidden shift corresponds to a possible isomorphism (whereas
there are many subgroups of either $S_{2n}$ or $S_n \wr \Z_2$ that do
not correspond to isomorphisms); and furthermore, viewed as black box
problems, the hidden shift problem can be reduced to the hidden
subgroup problem.  The hidden shift problem can be tackled on a
quantum computer using a standard method that closely parallels the
standard approach to the hidden subgroup problem.  We present two
hardness results for this standard approach to the hidden shift
problem over $S_n$.

First, we prove that $\Omega(n)$ copies of the hidden shift state are
necessary to solve the problem (whereas $O(n\log n)$ copies are
sufficient). The idea behind this bound is the simple observation that
the hidden shift problem for the largest abelian subgroup of $S_n$ is
at least as hard as for the whole group $S_n$. In the case where the
group $G$ is abelian, the hidden shift problem for $G$ is equivalent
to the hidden subgroup problem over the generalized dihedral group $G
\semidirect \Z_2$, and it is straightforward to obtain a reasonably
tight bound for this case using a connection to the subset sum problem
over $G$.  Since $S_n$ contains large abelian subgroups, the resulting
bound for the nonabelian hidden shift problem is not too bad.

Second, we present a simple proof that single-register measurements
are not sufficient to solve the hidden shift problem over $S_n$.  In
fact, this result holds for any group that has many irreducible
representations of sufficiently high degree.  In particular, the only
property of $S_n$ used in the proof is the fact that under the
Plancherel distribution, an irreducible representation of $S_n$ with
degree larger than $n^{\Theta(n)}$ occurs with probability at least
$1-n^{-\Omega(n)}$.

The remainder of the paper is organized as follows. In
Section~\ref{sec:shift} we define the nonabelian hidden shift problem
and discuss the standard approach to solving it.  In
Section~\ref{sec:iso} we discuss how isomorphism problems (including,
but not limited to, graph isomorphism) can be cast as hidden shift
problems.  In Section~\ref{sec:numcopies} we prove the linear lower
bound on the required number of copies of hidden shift states.  In
Section~\ref{sec:structure} we examine the structure of the hidden
shift states for arbitrary groups and obtain some results needed for
Section~\ref{sec:single}, where we show that single register
measurements are insufficient.  Finally, in the Appendix, we present
some additional results on the rank of the hidden shift states.

\section{Nonabelian hidden shift problem}
\label{sec:shift}

The (nonabelian) hidden shift problem is the following.  We are given
black-box access to two functions $f_0: G \to S$ and $f_1:G \to S$
where $G$ is a (nonabelian) group and $S$ is a finite set.  The
functions are promised to satisfy two conditions:
\begin{enumerate}
\item Both $f_0$ and $f_1$ are injective.
\item Either there exists a fixed hidden shift $s \in G$ such that
$f_0(g)=f_1(gs)$ for all $g \in G$, or the images of $f_0$ and $f_1$
are disjoint (in which case we say there is no hidden shift).
\end{enumerate}
The goal is to determine whether there is a hidden shift $s$ or not.

The case where $G$ is an abelian group has received considerable
attention \cites{EH00,Reg02,FIMSS03,DHI03,Kup03,Reg04,BCD05a}.  Since
inversion is an automorphsim of any abelian group, the hidden shift
problem in $G$ is a hidden subgroup problem in the generalized
dihedral group $G \semidirect \Z_2$ where $\Z_2$ acts by inversion.
In particular, the case where $G$ is cyclic is the well-known dihedral
hidden subgroup problem.  However, the case where $G$ is nonabelian,
in which case the hidden shift problem is not a hidden subgroup
problem, seems not to have been studied extensively.

In this paper, we focus on a particular natural approach to solving
the hidden shift problem on a quantum computer, paralleling the
standard quantum approach to the hidden subgroup problem.  First
prepare a uniform superposition over $i \in \Z_2$ and $g \in G$, and
then compute the value of $f_i(g)$, giving the state
\be
  \frac{1}{\sqrt{2 |G|}} \sum_{g \in G} 
  \big(|0,g,f_0(g)\> + |1,g,f_1(g)\>\big)
\,.
\ee
Then measure the third register.  If there is a hidden shift $s$, then
we are left with the state
\be
  |\phi_{s,g}\> := \frac{1}{\sqrt 2} (|0,g\> + |1,gs\>)
  \label{eq:hiddenShifts}
\ee
for some uniformly random (unknown) $g \in G$.  On the other hand, if
there is no hidden shift, we obtain the state $|i,g\>$ for some
uniformly random (unknown) $i \in \Z_2$ and $g \in G$.  
Thus the density matrix obtained by applying the procedure is
either
\ba
  \gamma_1(s) &:= \frac{1}{|G|} \sum_{g \in G} 
                   |\phi_{s,g}\>\<\phi_{s,g}|
\intertext{if there is a hidden shift $s \in G$, or the maximally
mixed state}
  \gamma_2     &:= \frac{1}{2|G|} \onemat_{2|G|}
\ea
if there is no hidden shift.  Using the state thus obtained, we would
like to decide whether there is a hidden shift or not.

In general, we can apply the above procedure $k$ times to obtain $k$
copies of the hidden shift state (or the maximally mixed state if
there is no hidden shift).  Clearly, these states become more
distinguishable as $k$ is increased.  Suppose that in the case where
there is a hidden shift $s$, it is equally likely to correspond to any
element of $G$.  Then the problem is to distinguish the two density
operators
\ba
  \gamma_1^{(k)} &:= 
    \frac{1}{|G|} \sum_{s \in G} \gamma_1^{(k)}(s)
    \label{eq:hiddenShift}\\
  \gamma_2^{(k)} &:= \frac{1}{(2 |G|)^k}\onemat
    \label{eq:noShift}
\,,
\ea
where $\gamma_1^{(k)}(s):=\gamma_1(s)^{\otimes k}$.

A natural generalization of the nonabelian hidden shift problem
involves the case of $M$ injective functions, $f_j$ for $j \in
\{0,1,\ldots,M-1\}$, satisfying $f_j(g)=f_{j+1}(gs)$ for a fixed $s \in
G$ for all $j \in \{0,1,\ldots,M-2\}$.  This problem becomes easier as
$M$ is increased, and is interesting in the case where $G$ is cyclic,
since it has an efficient quantum algorithm provided $M$ is
sufficiently large \cite{CD05}.  We will not consider the generalized
nonabelian hidden shift problem further in this paper, although it is
an interesting question whether this problem has an efficient quantum
algorithm even for $M$ sufficiently large.

\section{Isomorphism problems}
\label{sec:iso}

The nonabelian hidden shift problem for the symmetric group is
especially interesting since an efficient quantum algorithm for this
problem would yield an efficient algorithm for graph isomorphism (and
more generally, for other related isomorphism problems).  The usual
quantum approach to graph isomorphism relies on a reduction to the
hidden subgroup problem for the symmetric group, but the hidden shift
problem for $G=S_n$ presents an alternative approach that seems to be
at least as natural, and is arguably more so.

We now describe a generalized isomorphism problem that reduces to the
hidden shift problem.  For each $n \in \N$, let $\cC_n$ be a set of
objects of size $n$. For example, $\cC_n$ could be the set of graphs
on $n$ vertices.  We assume that the objects can be uniquely
represented using $\poly(n)$ bits.

Let $G_n$ be a family of (finite) groups such that each $G_n$ acts on
$\cC_n$. For $g \in G_n$ and $C\in \cC_n$, let $g(C)$ denote the
element of $\cC_n$ given by the action of $g$ on $C$.  We call two
objects $A,B\in\cC_n$ {\em isomorphic} if there is some $g \in G_n$
such that $g(A)=B$. We call an object $C\in\cC_n$ {\em rigid} if it
has no automorphisms, i.e., if there is no $g \in G_n \setminus \{1\}$
such that $g(C)=C$.

The {\em $\cC$-isomorphism problem} is the following.  Given two rigid
objects $C_0,C_1 \in \cC_n$, determine whether they are isomorphic or
nonisomorphic.  It is straightforward to reduce this isomorphism
problem to a corresponding hidden shift problem: simply let
$f_i(g):=g(C_i)$.  (The assumption of rigidity is required to ensure
that $f_0,f_1$ are injective.)

Graph isomorphism is the special case of the $\cC$-isomorphism problem
for $G_n=S_n$ where $\cC_n$ is the set of graphs on $n$ vertices, and
the action of $G_n$ is to permute the vertices.  Thus, a solution to
the generalized hidden shift problem for $G_n=S_n$ would give an
efficient algorithm for testing isomorphism of rigid graphs.  But such
an algorithm could also be applied to other isomorphism problems.  For
example, if we let $\cC_n$ be the set of all binary linear codes of
length $n$, where $G_n=S_n$ acts to permute the bits of the code
words, then we obtain the {\em code equivalence problem} discussed in
\cite{EH99}, which is at least as hard as graph isomorphism
\cite{PR97}.

As mentioned in the introduction, the usual approach to solving graph
isomorphism on a quantum computer is based not on the hidden shift
problem, but on the hidden subgroup problem.  Graph isomorphism can be
cast as a hidden subgroup problem over $S_{2n}$ where the hidden
subgroups are generated by full support involutions.  A more careful
inspection of the hidden subgroups that occur in this reduction shows
that it is sufficient work with a subgroup of $S_{2n}$: as proposed by
Ettinger and H{\o}yer, one can cast graph isomorphism as a
hidden subgroup problem over the wreath product $S_n \wr \Z_2 <
S_{2n}$ where the hidden subgroups are generated by so-called
involutive swaps \cite{EH99}.

How are the hidden subgroup and hidden shift approaches to graph
isomorphism related?  In general, one can show that the hidden shift
problem in a group $G$ can be reduced to the hidden subgroup problem
in $G \wr \Z_2$.  In particular, the hidden shift problem in $S_n$
reduces to the hidden subgroup problem in $S_n \wr \Z_2$ (and in fact,
using the results of \cite{WH05}, one can also show that it reduces to
the hidden subgroup problem in $S_{2n}$).  Since the hidden shift
problem is no harder than the corresponding hidden subgroup problem,
this suggests that the hidden shift problem might present a more
natural quantum approach to graph isomorphism.  However, we emphasize
that our hardness results about measurements of hidden shift states do
not imply corresponding results about hidden subgroup states, since
the reduction does not necessarily still hold when we assume the use
of the standard method to produce particular quantum states.

\section{Lower bound on the number of copies}
\label{sec:numcopies}

In this section, we show that $\Omega(n)$ copies of the hidden shift
states are needed to successfully determine whether there is a hidden
shift.  We do this by showing that the optimal POVM is unlikely to
produce the correct answer unless $k=\Omega(n)$.

Consider the general problem of distinguishing a pair of (possibly
mixed, a priori equiprobable) quantum states.  The optimal measurement
for this problem (in the sense that it maximizes the probability of
successfully identifying the state) was discovered by Helstrom
\cite{Hel76}, and is as follows.  Suppose we wish to distinguish the
quantum states $\rho_1,\rho_2$.  Then let $E_1$ be the projector onto
the eigenvectors of $\rho_1-\rho_2$ corresponding to positive
eigenvalues, and let $E_2$ be the projector onto the eigenvectors
of $\rho_1-\rho_2$ corresponding to negative eigenvalues.
(Eigenvectors in the nullspace of $\rho_1-\rho_2$ can be associated to
either $E_1$ or $E_2$ without affecting the success probability.)

In principle, Helstrom's result tells us the optimal measurement to
distinguish $\gamma_1^{(k)}$ and $\gamma_2^{(k)}$.  Unfortunately,
since we do not have a good understanding of the spectrum of
$\gamma_1^{(k)}$ for nonabelian groups, we do not know how to estimate
the success probability of the Helstrom measurement in such cases.
However, we can obtain a good estimate of the success probability for
abelian groups, and we can obtain a bound for arbitrary groups since a
bound for a subgroup implies a bound for the full group.
Specifically, we have

\begin{lemma}
\label{lem:subgroup}
The number of copies needed to solve the hidden shift problem in the
group $G$ (with a probability of success bounded above $1/2$ by a
constant) is at least as great as the number of copies needed to solve
the hidden shift problem in any subgroup $H \le G$.
\end{lemma}

\begin{proof}
Clearly, if the possible hidden shifts are restricted to be from a
subgroup $H \le G$, the problem is at least as hard as when the hidden
shift may be arbitrary.  For a uniformly random hidden shift $s \in
H$, the density matrix when there is a hidden shift is
\be
  \frac{1}{|H|} \sum_{h \in H} |\phi_{s,h}\>\<\phi_{s,h}|
\,,
\ee
which can be written as the tensor product of the unrestricted hidden
shift state in $H$ and a maximally mixed state of dimension $|G|/|H|$.
Since the maximally mixed state provides no information about the
hidden shift, the restricted problem in $G$ is equivalent to the
hidden shift problem for $H$. 
\end{proof}

Now we give a general lower bound on the number of copies needed to
solve an arbitrary abelian hidden shift problem.  In the abelian case,
we can give fairly tight bounds using the close connection between the
hidden shift problem and the subset sum problem \cites{BCD05a}.
Specifically, after performing a Fourier transform on the group
register, we can write the abelian hidden shift states as
\be
  \tilde \gamma_{1}^{(k)}(s) =
    \frac{1}{(2|G|)^k} \sum_{x \in G^k} \sum_{w,v \in G}
    \chi_w(s) \bar\chi_v(s) \sqrt{\eta^x_w \eta^x_v}
    |S^x_w,x\>\<S^x_v,x|
\ee
where
\be
  S^x_w := \{b \in \Z_2^k: b \cdot x = w\}
\ee
is the set of solutions of the subset sum problem over $G$, $\eta^x_w
:= |S^x_w|$ is the number of such solutions, and
\be
  |S^x_w\> := \frac{1}{\sqrt{\eta^x_w}} \sum_{b \in S^x_w} |b\>
\ee
is the normalized uniform superposition over those solutions (where we
define $|S^x_w\> := 0$ in the event that $\eta^x_w = 0$).  Thus, with
a uniformly random hidden shift, we have the state
\be
  \tilde\gamma_1^{(k)}
    = \frac{1}{(2|G|)^k} \sum_{x \in G^k} \sum_{w \in G}
      \eta^x_w |x,S^x_w\>\<x,S^x_w|
\,.
\ee

In the standard approach to the abelian hidden shift problem, our goal
is to distinguish this state from the maximally mixed state.  An
optimal measurement for doing so is the measurement that projects onto
the support of $\tilde\gamma_1^{(k)}$.  Since the eigenvalues of
$\tilde\gamma_1^{(k)}$ are integer multiples of $1/(2|G|)^k$, the
operator $\tilde\gamma_1^{(k)} - \tilde\gamma_2^{(k)}$ is nonnegative
precisely on the support of $\tilde\gamma_1^{(k)}$.  Therefore, the
projection onto that support is a Helstrom measurement, and hence is
optimal.

Having identified an optimal measurement, we can now show

\begin{lemma}
\label{lem:abelian}
For any abelian group $G$, $k=\Omega(\log |G|)$ copies of the hidden
shift states are needed to decide whether there is a hidden shift
(with a probability of success bounded above $1/2$ by a constant).
\end{lemma}

\begin{proof}
The success probability of the optimal measurement (in which $E_1$
projects onto the support of $\tilde\gamma_1^{(k)}$ and $E_2$ projects
onto its complement) is
\ba
  \Pr(\text{success})
    &:= \frac{1}{2} \Big(\tr E_1 \tilde\gamma_1^{(k)} 
                       + \tr E_2 \tilde\gamma_2^{(k)}\Big) \\
    &= 1 - \frac{\rank \tilde\gamma_1^{(k)}}{2(2|G|)^k}
\label{eq:prsuc}
\,.
\ea
Now
\ba
  \rank \tilde\gamma_1^{(k)}
    &= \sum_{x,w} \delta[\eta^x_w > 0] \\
    &= |G|^{k+1} - \sum_{x,w} \delta[\eta^x_w = 0]
\,.
\ea
(For the case $G=\Z_N$, the rank is given by the integer sequence
\cite{OEIS}*{A098966}.  For a discussion of the rank in the general
(not necessarily abelian) case, see the Appendix.)
To evaluate this expression, we need to understand the typical
behavior of $\eta^x_w$.  In particular, it is helpful to know the
first and second moments of $\eta^x_w$ for uniformly random $x \in
G^k$, $w \in G$.  For an arbitrary group $G$, the first moment is
\be
  \mu := \E_{x,w} \eta^x_w \\
      = \frac{2^k}{|G|}
\label{eq:mom1}
\,.
\ee
For the second moment, we have
\ba
  \E_{x,w} (\eta^x_w)^2
    &:= \frac{1}{|G|^{k+1}} \sum_{x,w} (\eta^x_w)^2 \\
    &= \frac{1}{|G|^{k+1}} \sum_{x,w} (\sum_b \delta_{b \cdot x, w})^2 \\
    &= \mu + \frac{1}{|G|^{k+1}} \sum_{x,w} \sum_{b \ne c} 
       \delta_{b \cdot x, c \cdot x} \, \delta_{b \cdot x, w} \\
    &= \mu + \frac{1}{|G|^{k+1}} \sum_x \sum_{b \ne c} 
       \delta_{b \cdot x, c \cdot x} \\
    &= \mu + \frac{2^k(2^k-1)}{|G|^2}
\label{eq:mom2}
\,.
\ea
Here in the final step we used the fact that for fixed $b \ne c$ (with
$b_k \ne c_k$ without loss of generality), and for fixed
$x_1,\ldots,x_{k-1} \in G$, there is exactly one $x_k \in G$ such that
$b \cdot x = c \cdot x$.
In terms of the variance
$
  \sigma^2 := \E_{x,w} (\eta^x_w)^2 - \mu^2
$
we have the inequality
$
  \Pr(\eta^x_w = 0) \le \sigma^2/(\mu^2 + \sigma^2)
$
\cite{AS00}, giving
\ba
  \rank \gamma_1^{(k)} 
    &\ge |G|^{k+1} \frac{\mu^2}{\mu^2 + \sigma^2} \\
    &=   |G|^{k+1} \bigg(\mu + 1 - \frac{1}{|G|}\bigg)^{-1} \\
    &\ge (2|G|)^k - |G|^{k+1}
\,.
\ea
Thus, we find
\be
  \Pr(\text{success})
    \le \frac{1}{2} \bigg(1 + \frac{|G|}{2^k}\bigg)
\,.
\ee
For the success probability to be bounded above $1/2$ by a
constant, we need $k=\Omega(\log |G|)$ as claimed.
\end{proof}

Putting these lemmas together, we have

\begin{theorem}
To solve the hidden shift problem in $S_n$, $\Omega(n)$ copies of the
hidden shift states are necessary.
\end{theorem}

\begin{proof}
The largest abelian subgroup of $S_n$ has size $3^{\Theta(n)}$
\cite{BM67} (see also \cite{OEIS}*{A000792}).
Combining Lemmas~\ref{lem:subgroup} and \ref{lem:abelian} gives the
result.
\end{proof}

This result is not too far from the best possible, since $O(\log |G|)$
copies are sufficient to solve the hidden shift problem for any group
$G$.  This follows easily from (\ref{eq:prsuc}) and the fact that
$\rank \tilde\gamma^{(k)}_1 \le |G|^{k+1}$, and is analogous to the
well-known result that $O(\log |G|)$ copies of hidden subgroup states
are sufficient to solve the hidden subgroup problem \cite{EHK04}.
However, there is a logarithmic gap between these lower and upper
bounds.  We suspect that the lower bound could be improved, since it
only uses information about abelian subgroups, but without a better
understanding of the structure of the hidden shift states for large
$k$, it seems difficult to establish a bound.

Note that an analogous bound of $\Omega(n)$ has recently been
independently established for the hidden subgroup problem over
$S_{2n}$ where the hidden subgroup may be an arbitrary full-support
involution, and a bound of $\Omega(n \log n)$ has been established for
the hidden subgroup problem over $S_n \wr \Z_2$ \cite{HRS05}.

It is worth noting that while the projection onto the support of
$\gamma^{(k)}_1$ is an optimal measurement in the abelian case, it is
not an optimal measurement in general.  For example, for
$G=S_4$, $\gamma^{(3)}_1$ has eigenvalues between $0$ and
$1/(2|G|)^3$, so that the projection onto the support is not a
Helstrom measurement.

\section{Structure of hidden shift states}
\label{sec:structure}

To show that single-register measurements are not sufficient to solve
the hidden shift problem, we need to understand the structure of the
states $\gamma_1^{(k)}(s)$, $\gamma_1^{(k)}$, and $\gamma_2^{(k)}$.
Here we determine their block structure and use it to compute the
spectrum of $\gamma_1^{(k)}$ for $k=1$ and $2$.

Observe that $\gamma_1(s)$ has the following form:
\ba
  \gamma_1(s) &= \frac{1}{2|G|} \sum_{g\in G}
  |0,g\>\<0,g| +
  |1,gs\>\<1,gs| +
  |0,g\>\<1,gs| +
  |1,gs\>\<0,g| \\
  &=
  \frac{1}{2|G|}
  \begin{pmatrix}
  I & R(s) \\
  R(s^{-1}) & I
  \end{pmatrix}
\,,
\ea
where $R$ is the {\em right regular representation} of $G$, defined by
\be
  R(s) |g\> = |gs^{-1}\>
\ee
for all $s,g \in G$. Recall that the regular representation contains
all irreducible representations of $G$ with multiplicities given by
their dimensions.  More precisely, we have
\be
  F \,R(s)\, F^\dagger = \bigoplus_{\rho\in\hat{G}}
                         I_{d_\rho} \otimes \rho(s)
\ee
for all $s \in G$, where $F$ is the Fourier transform over $G$ and
$\hat{G}$ is a complete set of irreducible representations of $G$.  In
other words, the Fourier transform decomposes the regular
representation into its irreducible constituents.

Using the Fourier transform, the states $\gamma_1^{(k)}(s)$,
$\gamma_1^{(k)}$, and $\gamma_2^{(k)}$ can be simultaneously block
diagonalized for any $k \in \N$. The blocks are enumerated by
$k$-tuples of irreducible representations.  In particular, in the
Fourier basis we have
\ba
  \tgamma_1^{(k)}(s) &= \frac{1}{(2|G|)^k}
    \bigoplus_{(\rho_1,\ldots,\rho_k)\in\hat{G}^k} \onemat_{d_{\rho_1}
    \cdots d_{\rho_k}} \otimes B^{\rho_1,\ldots,\rho_k}(s) \\
  \tgamma_1^{(k)} &= \frac{1}{(2|G|)^k}
    \bigoplus_{(\rho_1,\ldots,\rho_k)\in\hat{G}^k} \onemat_{d_{\rho_1}
    \cdots d_{\rho_k}} \otimes B^{\rho_1,\ldots,\rho_k} \\ 
  \tgamma_2^{(k)} &= \frac{1}{(2|G|)^k} 
    \bigoplus_{(\rho_1,\ldots,\rho_k)\in\hat{G}^k} \onemat_{d_{\rho_1}
    \cdots d_{\rho_k}} \otimes \onemat_{2 d_{\rho_1} \cdots 2 d_{\rho_k}}
\ea
where
\ba
  B^{\rho_1,\ldots,\rho_k}(s) &:= \bigotimes_{j=1}^k \left(
    \begin{array}{cc}
      \onemat_{d_{\rho_j}} & \rho_j(s) \\
      \rho_j(s^{-1})            & \onemat_{d_{\rho_j}}
    \end{array}
    \right) \\
  B^{\rho_1,\ldots,\rho_k} &:=
    \frac{1}{|G|}\sum_{s\in G} B^{\rho_1,\ldots,\rho_k}(s)
\,.
\ea
Here the factor $d_{\rho_1}\cdots d_{\rho_k}$ accounts for the
multiplicity of $(\rho_1,\ldots,\rho_k)$ in $k$ copies of the regular
representation of $G$.

It is straightforward to check that the blocks
$B^{\rho_1,\ldots,\rho_k}(s)$ and $B^{\rho_1,\ldots,\rho_k}$ can be
expressed as
\ba
  B^{\rho_1,\ldots,\rho_k}(s)
  &= \sum_{x,y\in\{0,1\}^k} \ket{x}\bra{y}
     \otimes
     A^{\rho_1,\ldots,\rho_k}_{y_1-x_1,\ldots,y_k-x_k}(s) \\
     B^{\rho_1,\ldots,\rho_k}
  &= \sum_{x,y\in\{0,1\}^k} \ket{x}\bra{y}
     \otimes
     A^{\rho_1,\ldots,\rho_k}_{y_1-x_1,\ldots,y_k-x_k}\,,
\ea
where
\ba
  A^{\rho_1,\ldots,\rho_k}_{z_1,\ldots,z_k}(s) &:=
    \rho_1(s^{z_1})\otimes \rho_2(s^{z_2}) \otimes \cdots \otimes
    \rho_k(s^{z_k}) \\
  A^{\rho_1,\ldots,\rho_k}_{z_1,\ldots,z_k} &:=
    \frac{1}{|G|} \sum_{s\in G} A^{\rho_1,\ldots,\rho_k}_{z_1,\ldots,z_k}(s)
\label{eq:blocksum}
\ea
for all $z\in\{-1,0,1\}^k$. Clearly, the matrices
$A^{\rho_1,\ldots,\rho_k}_{z_1,\ldots,z_k}$ are hermitian, that is,
$A^{\rho_1,\ldots,\rho_k}_{z_1,\ldots,z_k}=
A^{\rho_1,\ldots,\rho_k}_{-z_1,\ldots,-z_k}$.

To understand the form of these matrices, we must carry out the sum in
(\ref{eq:blocksum}) for various choices of the irreducible
representations $\rho_1,\ldots,\rho_k \in \hat{G}$ and the indices
$z_1,\ldots,z_k \in \{-1,0,1\}$.  If all $z_j$ have the same sign,
then such a calculation is straightforward, using the following
well-known result:

\begin{lemma}\label{lem:averageRep}
Let $\pi$ be any representation of the group $G$. Then the
matrix
\be
A:=\frac{1}{|G|} \sum_{g\in G} \pi(g)
\ee
is a projection operator whose rank is the number of times the trivial
representation appears in $\pi$.
\end{lemma}

\begin{proof}
Decompose the representation $\pi$ into irreducible
representations. Let $\sigma$ be any irreducible representation
occurring in $\pi$. The sum $B=\frac{1}{|G|} \sum_{g\in G} \sigma(g)$
is a multiple of the identity matrix because $B$ commutes with all
$\sigma(h)$ for $h\in G$. The trace of $B$ is the inner product of the
trivial character and the character of $\sigma$. Therefore, $B=1$ if
$\sigma$ is the trivial representation and $B$ is the zero matrix if
$\sigma$ is not the trivial representation.
\end{proof}

In general, we will have $z_j$'s of both signs.  In this case we may
say that $A$ includes both representations and antirepresentations of
$G$, since $g \mapsto \rho(g^{-1})$ is a group antihomomorphism.
Fortunately, this case can be dealt with using the following:

\begin{lemma}\label{lem:averageRepAnti}
Let $\rho$ and $\sigma$ be two irreducible representations of $G$.
Then the entries of the matrix
\be
  A:=\frac{1}{|G|} \sum_{g\in G} \rho(g)\otimes \sigma(g^{-1})
\ee
are given by
\be
  A_{i,j;k,l} =
  \delta_{\rho,\sigma}\, \frac{1}{d_\rho}\, \delta_{i,l}\, \delta_{j,k}
  \label{eq:twist}
\ee
where $i,j$ are the row and column indices of the first tensor
component and $k,l$ are the row and column indices of the second
tensor component.
\end{lemma}

\begin{proof}
The entries are given by
\be
  A_{i,j;k,l} =
  \frac{1}{|G|} \sum_{g\in G}
  \rho_{ij}(g)\otimes \overline{\sigma_{lk}(g)} \\
\,;
\ee
then (\ref{eq:twist}) follows directly from the Schur orthogonality
relations.
\end{proof}

Now we are ready to investigate the blocks $B^\rho$ for $k=1$ and the
blocks $B^{\rho,\rho}$ for $k=2$.

\begin{lemma}[Spectrum for $k=1$]
\label{lem:spect1}
The block $B^\triv$ has eigenvalues $2$ and $0$.  For $\rho \ne
\triv$, $B^\rho = I_{2 d_\rho}$.
\end{lemma}

\begin{proof}
Since $\rho_\triv(s)=1$ for all $s$,
\be
  B^\triv = \begin{pmatrix}1 & 1 \\ 1 & 1 \end{pmatrix}
\,,
\ee
which has eigenvalues $2$ and $0$.  For $\rho \ne \triv$, $\sum_{s \in
G} \rho(s) = 0$ by the orthogonality of $\rho$ and $\triv$, so that
$B^\rho = \onemat_{2 d_\rho}$ as claimed.
\end{proof}

\begin{lemma}[Spectrum for $k=2$]
\label{lem:spect2}
For any $\rho \in \hat G \setminus \{\triv\}$, either $B^{\rho,\rho}$
has the spectrum $1$ (multiplicity $2 d_\rho^2$) and $1 \pm 1/d_\rho$
(multiplicity $d_\rho^2$ each); or the spectrum $2$ (multiplicity 1),
$0$ (multiplicity 1), $1$ (multiplicity $2 d_\rho^2
- 2$), and $1 \pm 1/d_\rho$ (multiplicity $d_\rho^2$ each).
\end{lemma}

\begin{proof}
For simplicity, we omit the label $\rho,\rho$.  The block of
interest has the form
\be\label{eq:M2}
  B=
  \begin{pmatrix}
    \onemat   & A_{ 0, 1} & A_{ 1, 0} & A_{ 1, 1} \\
    A_{ 0, 1} & \onemat   & A_{ 1,-1} & A_{ 1, 0} \\
    A_{ 1, 0} & A_{ 1,-1} & \onemat   & A_{ 0, 1} \\
    A_{ 1, 1} & A_{ 1, 0} & A_{ 0, 1} & \onemat      
  \end{pmatrix}
\,.
\ee
Recall that the blocks of $B$ are enumerated by $x,y\in\{0,1\}^2$. The
matrix at position $(x,y)$ is given by $A_{y-x}$, where the
$A$ matrices are defined by
\be
  A_z :=
  \frac{1}{|G|}\sum_{g\in G}
  \rho(g^{z_1}) \otimes \rho(g^{z_2})
\ee
for $z\in\{-1,0,1\}^2$.  We have simplified (\ref{eq:M2}) to minimize
the number of $-1$'s using the fact that $A_z$ is hermitian, so
$A_z=A_{-z}$.

Since $\rho \ne \triv$ by assumption, $A_{0,1}=A_{1,0}=0$ by the
calculation in Lemma~\ref{lem:spect1}.  Thus
\be
  B=
  \begin{pmatrix}
    \onemat & 0        & 0        & A_{1,1} \\
    0       & \onemat  & A_{1,-1} & 0 \\
    0       & A_{1,-1} & \onemat  & 0 \\
    A_{1,1} & 0        & 0        & \onemat
  \end{pmatrix}
  \cong
  \begin{pmatrix}
    \onemat       & A_{1,1} \\       
    A_{1,1}       & \onemat
  \end{pmatrix}
  \oplus
  \begin{pmatrix}
    \onemat       & A_{1,-1} \\       
    A_{1,-1}      & \onemat
  \end{pmatrix}
\,.
\ee
Hence it remains to understand the operators $A_{1,1}$ and $A_{1,-1}$.

Since $\rho$ (and hence also $\bar\rho$) is irreducible, the trivial
representation appears at most once in $\rho \otimes \rho$, so by
Lemma~\ref{lem:averageRep}, $A_{1,1}$ is either zero or a projector of
rank one.  Hence the matrix
\be
  \begin{pmatrix} I & A_{1,1} \\ A_{1,1} & I \end{pmatrix}
\ee
is either the identity, or has the eigenvalues $2$ and $0$ with
multiplicity $1$, and $1$ with multiplicity $2 d_\rho^2 - 1$.
By Lemma~\ref{lem:averageRepAnti}, $A_{1,-1}$ has eigenvalues $\pm
1/d_\rho$, so that
\be
  \begin{pmatrix} I & A_{1,-1} \\ A_{1,-1} & I \end{pmatrix}
\ee
has the eigenvalues $1 \pm 1/d_\rho$ each with multiplicity $d_\rho^2$.
\end{proof}

\section{Single-register measurements do not suffice}
\label{sec:single}

In this section we show that single-register measurements do not
suffice to efficiently solve the hidden shift problem for $G=S_n$.

Let us first explain in more detail what is meant by an algorithm
restricted to single-register measurements.  A POVM $\cE$ with a set
of possible outcomes $J$ is a collection of positive operators
$\cE=\{E_j: j\in J\}$ satisfying the completeness condition
\be
  \sum_j E_j = \onemat
\,.
\ee
An efficient algorithm consists of a polynomial number of POVMs
$\cE_1,\ldots,\cE_t$, each acting on a single copy of the hidden shift
state.  After obtaining the measurement outcomes $j_1,\ldots,j_t$, a
final computation is performed to decide whether there is a hidden
shift or not.  Note that the individual outcomes $j_i$ need not
directly correspond to one situation or the other.  Also, let us
stress out that the POVMs $\cE_1,\ldots,\cE_t$ may be chosen
adaptively, that is, $\cE_r$ may depend on all previous outcomes
$j_1,\ldots,j_{r-1}$ for $2\le r\le t$.

To simplify the analysis, we can refine any POVM $\cE$ so that each
$E_j=a_j\ket{\psi_j}\bra{\psi_j}$ where each $\ket{\psi_j}$ is a unit
vector and $a_j > 0$ without loss of generality.  This is because
any positive operator can be written as a weighted sum of projection
operators, where the weights correspond to the eigenvalues and the
projection operators to the eigenspaces. The result of this
measurement on the state $\gamma$ is a random variable, where we
obtain $j\in J$ with probability
\be
  p(j) = a_j\bra{\psi_j}\gamma\ket{\psi_j}
\,.
\ee

In our case, the POVM can be further simplified because the states
$\gamma_{1}^{(k)}(s)$, $\gamma_1^{(k)}$, and $\gamma_2^{(k)}$ can be
simultaneously block-diagonalized as described in
Section~\ref{sec:structure}.  The blocks are labeled by irreducible
representations of $G$. Therefore, as in the hidden subgroup problem,
we may assume without loss of generality that we first perform a
Fourier transform on the group register and then measure the
representation name (so-called {\em weak Fourier sampling}).  Next, we
perform a measurement within the subspace corresponding to the
observed representation.

{From} the block decomposition of the states described in
Section~\ref{sec:structure}, it is clear that the
various irreducible representations of $G$ occur independently
according to the Plancherel distribution, i.e.,
\be
  \Pr(\rho) = \frac{d_{\rho}^2}{|G|}
\,,
\ee
regardless of whether or not there is a hidden shift.  This is
analogous to the fact that weak Fourier sampling is insufficient to
distinguish between the trivial subgroup and the subgroups generated
by full support involutions in the symmetric group \cite{HRT03}.

Suppose we measure the representation name and observe a particular
$\rho\in\hat{G}$.  Then consider an arbitrary POVM
$\cE=\{a_1\ket{\psi_1}\bra{\psi_1},\ldots,a_r\ket{\psi_r}\bra{\psi_r}\}$
acting on the subspace of dimension $2d_\rho$ corresponding to the
observed representation.

If there is no hidden shift (that is, if the state is
$\gamma_2^{(1)}$), then the post-measurement state is
$\onemat_{2d_\rho}/(2 d_\rho)$, and the probability of obtaining the
outcome $j$ is
\be
  p_2(j) = \frac{a_j}{2d_\rho} \bra{\psi_j} \onemat_{2d_\rho} \ket{\psi_j} 
         = \frac{a_j}{2d_\rho}
\,.
\ee
We denote this probability distribution by $P_2$.  On the other hand,
if there is a hidden shift $s$, then the post-measurement state is
$B^{\rho}(s)/(2 d_\rho)$, and the probability of obtaining the outcome
$j$ is
\be
  p_1(j|s) := \frac{a_j}{2d_\rho}
              \bra{\psi_j} B^\rho(s) \ket{\psi_j}
\,;
\ee
we denote this distribution by $P_{1,s}$.  We will also be interested
in the distribution $P_1$ obtained by averaging over $s \in G$, i.e.,
with the probabilities
\be
  p_1(j) := \frac{1}{|G|} \sum_{s\in G} p_1(j|s)
\,.
\ee

Following \cites{MRS05,MR05}, the strategy for proving that
single-register measurements are not sufficient is to show that with
high probability (over the hidden shift $s$ and the observed
representation $\rho$), the statistics of the measurement results when
there is a hidden shift $s$ are close to those when there is no hidden
shift.  More precisely, we will prove

\begin{theorem}
\label{thm:single}
\be
  \Pr_{s \in G, \rho \in \hat G}
  ( \norm{P_{1,s} - P_2}_1 \ge e^{-\Theta(n)} ) 
  \le e^{-\Theta(n)}
\ee
\end{theorem}

To prove this theorem, we first show that with high probability (over
a uniformly random choice of $s \in G$ and the Plancherel distribution
of irreducible representations $\rho \in \hat{G}$), the distribution
$P_{1,s}$ is close to the distribution $P_1$.  Then it suffices to
show that $P_1$ and $P_2$ are typically close, which is
straightforward (since in fact, they are typically identical).

Because $P_1$ is the average of $P_{1,s}$ over $s \in G$, we can show
that the distributions are likely to be close by showing that the
variance of $p_1(j|s)$ is small (so that we can apply the Chebyshev
inequality).  More precisely, we will use the following:

\begin{lemma}[Upper bound on the sum of weighted variances]
Assume we have measured the irreducible representation $\rho \ne
\triv$, and we perform an arbitrary measurement
$\cE=\{a_j\ket{\psi_j}\bra{\psi_j}: j \in J\}$.  Then
\be
  \sum_{j\in J} \frac{\sigma_j^2}{a_j} \le \frac{1}{d_\rho^2}
\ee
where $\sigma^2_j$ is the variance of $p_1(j|s)$ when $s$ is chosen
uniformly from $G$.
\label{lem:variance}
\end{lemma}

\begin{proof}
For any fixed $j$ the variance $\sigma_j^2$ is given by
\be
  \sigma_j^2:=\frac{1}{|G|} \sum_{s\in G} p_1(j|s)^2 - p_1(j)^2\,.
\ee
Recall that we have $p_1(j)=a_j/(2d_\rho)$ for all $j$. This is
because we have $B^\rho=I_{2d_\rho}$ for all $\rho \ne \triv$ as shown
in Lemma~\ref{lem:spect1}.

The second moment can be expressed in terms of the block
$B^{\rho,\rho}$. We have
\ba
  \frac{1}{|G|} \sum_{s\in G} p_1(j|s)^2
    &=  \frac{a_j^2}{(2d_\rho)^2} \frac{1}{|G|}
        \sum_{s\in G} \big(\<\psi_j|B^\rho(s)|\psi_j\>\big)^2 \\
    &=  \frac{a_j^2}{(2d_\rho)^2} \frac{1}{|G|}
        \sum_{s\in G} \<\psi_j|\<\psi_j|
        B^{\rho}(s)\otimes B^{\rho}(s) |\psi_j\>|\psi_j\>\\
    &=  \frac{a_j^2}{(2d_\rho)^2} \frac{1}{|G|}
        \sum_{s\in G} \<\psi_j|\<\psi_j|B^{\rho,\rho}(s)|\psi_j\>|\psi_j\>\\
    &=  \frac{a_j^2}{(2d_\rho)^2}
        \<\psi_j|\<\psi_j|B^{\rho,\rho}|\psi_j\>|\psi_j\>
\,.
\ea
Set $\Delta:=|B^{\rho,\rho}-I|$.  Then we have for the variance the
upper bound
\be
  \sigma_j^2 \le
  \frac{a_j^2}{(2d_\rho)^2}
  \<\psi_j|\<\psi_j| \Delta |\psi_j\>|\psi_j\>
\,.
\ee
The operator $\Delta$ has the eigenvalue $1$ occurring with
multiplicity either $0$ or $2$ and the eigenvalue $1/d_\rho$ occurring
with multiplicity $2d_\rho^2$.  This follows from
Lemma~\ref{lem:spect2} where we have determined the spectrum of blocks
of the form $B^{\rho,\rho}$.  Denote the spectral decomposition of
$\Delta$ by
\be
  \Delta = P + \frac{1}{d_\rho} Q
\ee
where $P,Q$ are projectors.  We bound the sum of the weighted
variances by looking at $P$ and $Q/d_\rho$ separately. We have
\be
      \sum_{j\in J} a_j \<\psi_j|\<\psi_j| Q/d_\rho |\psi_j\>|\psi_j\>
  \le \sum_{j\in J} \frac{a_j}{d_\rho} 
    = 2
\,.
\ee
We also have
\be
  \sum_{j\in J} a_j \<\psi_j|\<\psi_j| P |\psi_j\>|\psi_j\> \le
  \rank P \le 2
\ee
where the first inequality follows by Lemma~12 in \cite{MRS05}.
Putting these two bounds together and multiplying by $1/(2d_\rho)^2$,
we obtain the desired result.
\end{proof}

Now we can use this result to show that $P_{1,s}$ and $P_1$ are
probably close:

\begin{lemma}
\label{lem:closetoavg}
\be
  \Pr_{s \in G, \rho \in \hat G}
  ( \norm{P_{1,s} - P_1}_1 \ge e^{-\Theta(n)} )
  \le e^{-\Theta(n)}
\ee
\end{lemma}

\begin{proof}
For any fixed representation $\rho \in \hat G$, according to
Chebyshev's inequality,
\be
  \Pr_{s \in G} \big(\big|p_1(j|s) - p_1(j)\big| \ge a_j c\big)
    \le \frac{\sigma_j^2}{a_j^2 c^2}
\ee
for any $c > 0$.  Now define
\be
  J_{\rm bad}^s := \big\{j \in J: 
                         \big|p_1(j|s)-p_1(j)\big| \ge a_j c\big\}
\,,
\ee
and define $J_{\rm good}^s := J \setminus J_{\rm bad}^s$.  The total
variation distance can be decomposed into contributions from good and
bad $j$'s.  For the good $j$'s, we have
\ba
  \sum_{j \in J_{\rm good}^s} \big|p_1(j|s)-p_1(j)\big|
    &\le \sum_{j \in J_{\rm good}^s} a_j c \\
    &\le 2 d_\rho c
\,.
\ea
Now for any $j \in J$ (and in particular, for $j \in J_{\rm bad}^s$),
we have
\ba
  \big|p_1(j|s) - p_1(j)\big|
    &=   \frac{a_j}{2 d_\rho}
         \big|\<\psi_j|B^\rho(s) - B^\rho|\psi_j\>\big| \\
    &\le \frac{a_j}{2 d_\rho}
         \norm{B^\rho(s) - B^\rho} \\
    &\le \frac{a_j}{d_\rho}
\,.
\ea
Thus it suffices to show that $\sum_{j \in J_{\rm bad}^s} a_j$ is
small.  The expectation of this quantity is
\ba
  \E_{s \in G} \sum_{j \in J_{\rm bad}^s} a_j
    &=   \frac{1}{|G|} \sum_{s \in G} \sum_{j \in J} 
         a_j \, \delta[j \in J^s_{\rm bad}] \\
    &=   \sum_{j \in J} a_j \Pr_{s \in G}(j \in J^s_{\rm bad}) \\
    &\le \sum_{j \in J}
         \frac{\sigma_j^2}{a_j c^2} \\
    &\le \frac{1}{d_\rho^2 c^2}
\ea
where in the last line we have used Lemma~\ref{lem:variance} (assuming
$\rho \ne \triv$, which we will later ensure).  Hence by Markov's
inequality,
\be
  \Pr\bigg( \sum_{j \in J_{\rm bad}^s} a_j \ge c' \bigg) 
    \le \frac{1}{d_\rho^2 c^2 c'}
\ee
for any $c' > 0$.  Conditioning on this event, we have
\ba
  \norm{P_{1,s} - P_1}_1
    &=   \sum_{j \in J_{\rm good}^s} \big|p_1(j|s)-p_1(j)\big|
        +\sum_{j \in J_{\rm bad}^s} \big|p_1(j|s)-p_1(j)\big| \\
    &\le 2 d_\rho c
        +\frac{c'}{d_\rho}
\ea
with probability at least
\be
  1-\frac{1}{d_\rho^2 c^2 c'}
\,.
\ee
Hence if we choose
\ba
  c  &= \frac{e^{-\alpha n}}{d_\rho} \\
  c' &= e^{3 \alpha n}
\ea
for some fixed $\alpha > 0$, we find
\be
  \norm{P_{1,s} - P_1}_1 \le 2 e^{-\alpha n} + \frac{e^{3 \alpha n}}{d_\rho}
\label{eq:distbound}
\ee
with probability at least
\be
  1 - e^{-\alpha n}
\,.
\ee
For $P_{1,s}$ and $P_1$ to be close with high probability, it
suffices that $d_\rho$ is large with high probability, so that the
second term of (\ref{eq:distbound}) is small.  Thus we condition on
the event that $d_\rho > n^{c'' n}$ for some constant $c''$, which
occurs with probability at least $1-n^{-\Omega(n)}$
\cite{MRS05}*{Lemma 6}.  This completes the proof.
\end{proof}

Finally, we must show that the probability distributions $P_1$ and $P_2$
are close in total variation distance:

\begin{lemma}
\label{lem:avgclosetonatural}
For an arbitrary POVM
acting on a single copy of the hidden shift state,
\be
  \| P^{\rho}_1 - P^{\rho}_2 \|= 0
\ee
for $\rho\neq\triv$ and
\be
  \| P^{\triv}_1 - P^{\triv}_2 \| \le \frac{1}{2}
\ee
for the trivial representation $\triv$.
\end{lemma}

\begin{proof}
Let $B$ be the block corresponding to the measured representation. Let
$\Delta:=|I-B|$.  Then we have
\ba
  \|P_1-P_2\| 
    &=   \frac{1}{2d_\rho} \sum_j a_j 
         |\bra{\psi_j}I_{d_\rho}\ket{\psi_j} 
        - \bra{\psi_j}B\ket{\psi_j}| \\
    &\le \frac{1}{2d_\rho} \sum_j a_j 
         \bra{\psi_j}\Delta\ket{\psi_j} \\
    &=   \frac{1}{2d_\rho} \sum_j \tr(a_j\ket{\psi_j}\bra{\psi_j}\Delta) \\
    &=   \frac{1}{2d_\rho} \tr(\Delta)
\,.
\ea
We have determined the spectrum of $B$ in Lemma~\ref{lem:spect1},
from which the lemma follows.
\end{proof}

Putting these results together, we can now prove the main result:

\begin{proof}[Proof of Theorem~\ref{thm:single}]
Since the trivial representation only appears with probability $1/n!$,
we can simply condition on not obtaining the trivial representation,
and the result follows from Lemmas~\ref{lem:closetoavg} and
\ref{lem:avgclosetonatural}.
\end{proof}

\section*{Acknowledgments}

We thank Dorit Aharonov, Sean Hallgren, Martin R{\"o}tteler, and
Pranab Sen for helpful discussions about the relationship between the
hidden shift and hidden subgroup problems.  We thank Sergey Bravyi for
a discussion about the decomposition of the product of a
representation and an antirepresentation.  And we thank David Wales
for discussions about the rank of $\gamma_1^{(k)}$, and in particular,
for correctly conjecturing the exact value of $\rank \gamma_1^{(2)}$
for an arbitrary group.
This work was supported in part by the National Science Foundation
under contract number PHY-0456720 and by the National Security Agency
under Army Research Office contract number W9111NF-05-1-0294.

\appendix
\section{Rank calculations}

Although the measurement that projects on the support of
$\gamma_1^{(k)}$ need not be optimal in general, it is nevertheless a
natural measurement to consider---for example, an analogous
measurement was used in \cite{EH99} to show that $O(n \log n)$ hidden
subgroup states are sufficient to solve a hidden subgroup problem
relevant to graph isomorphism.  Since we are trying to distinguish
$\gamma_1^{(k)}$ from the maximally mixed state, the success
probability of the measurement that projects onto the support depends
only on the rank of $\gamma_1^{(k)}$ (see (\ref{eq:prsuc})).  Here we
summarize some results on the rank for $k=1$ and $2$.

For the case $k=1$, Lemma~\ref{lem:spect1} immediately gives
\be
  \rank \gamma_1^{(1)} = 2|G|-1
\,.
\ee

For the case $k=2$, Lemma~\ref{lem:spect2} gives the contribution to
the rank from the cases where the same irreducible representation
$\rho \ne \triv$ occurs twice.  It is straightforward to calculate the
contribution from the other cases, giving the final result
\ba
  \rank \gamma_1^{(2)} 
    &= 4|G|^2 - 6 |G| + 3 + \sum_{\rho\in\hat{G},\,d_\rho>1} d_\rho^2 \\
    &= 4|G|^2 - 5 |G| + 3 - |\{\rho \in \hat G: d_\rho = 1\}|
\,.
\ea
In particular, for $G=S_n$, we have $|G|=n!$ and only two
one-dimensional representations (the trivial and sign
representations), so
\be
  \rank \gamma_1^{(2)} = 4 (n!)^2 - 5 n! + 1
\,.
\ee

Calculations of the rank for larger $k$ would seem to require a better
understanding of the structure of $\gamma_1^{(k)}$.


\end{document}